\documentclass[traditabstract]{aa} 

\usepackage{graphicx}
\usepackage{txfonts}
\usepackage{subfigure}
\usepackage{subfig}
\usepackage{amssymb}
\usepackage{natbib}
\bibpunct{(}{)}{;}{a}{}{,} 
\usepackage{url}

\begin{document} 

\titlerunning {First detection of   [N~{\sc II}] 205~$\mu$m absorption in interstellar gas}
   \title{First detection of   [N~{\sc II}] 205~$\mu$m  absorption in interstellar gas}

 \subtitle{{\emph{Herschel}}\thanks{\emph{Herschel} is an ESA space observatory with science instruments provided by European-led  Principal Investigator consortia and with important participation from NASA.}-HIFI  observations  towards  W31C, W49N, W51, and G34.3+0.1}
\authorrunning {C.M.~Persson, M.~Gerin, B.~Mookerjea, et.~al.}  
   \author{C.M.~Persson,     
          \inst{1},
          M.~Gerin\inst{2}, 
B.~Mookerjea\inst{3}, 
         J.H.~Black\inst{1}, 
M.~Olberg\inst{1}, 
J.R.~Goicoechea\inst{4},
G.E.~Hassel\inst{5}, 
E.~Falgarone\inst{2}, 
F.~Levrier\inst{2}, 
K.M.~Menten\inst{6},
J.~Pety\inst{7}
          }

   \institute{Chalmers University of Technology, Department of Earth and Space Sciences, Onsala Space Observatory,  SE-439 92 Onsala, Sweden.
    \email{\url{carina.persson@chalmers.se}} 
\and LERMA-LRA, UMR 8112 du CNRS, Observatoire de Paris, \'Ecole Normale
Sup\'erieure, UPMC \& UCP, 24 rue Lhomond, 75231 Paris Cedex 05, France 
\and   Tata Institute of Fundamental Research, Homi Bhabha Road, Mumbai 400005, India   
\and Instituto de Ciencia de Materiales de Madrid (ICMM-CSIC). E-28049, Cantoblanco, Madrid, Spain 
\and Department of Physics \& Astronomy, Siena College, Loudonville, NY  12211,   USA  
\and  Max-Planck-Institut f\"ur Radioastronomie, Auf dem H\"ugel 69, D-53121 Bonn, Germany  
\and Institut de Radioastronomie Millimétrique, 300 Rue de la Piscine, F-38406 Saint Martin d'Hères, France 
}

   \date{Received April 15, 2014; accepted June 11, 2014}

 
  \abstract
   {We present high resolution  [N~{\sc II}] 205~$\mu$m \mbox{($^3P_1 - ^3P_0$)} spectra obtained with 
\emph{Herschel}-HIFI   towards a small sample of
  far-infrared bright star forming regions in the Galactic plane: W31C (G10.6$-$0.4), W49N (G43.2$-$0.1), 
W51 (G49.5$-$0.4), and G34.3+0.1.
All sources display an  emission line profile    associated directly with the  H~{\sc II} ~regions 
themselves. 
For the first time  we also detect  \emph{absorption} of the [N~{\sc II}] 205~$\mu$m line   by extended 
low-density foreground material 
towards W31C and W49N 
over a wide range 
of velocities. We attribute this absorption to 
the warm ionised medium (WIM) and find 
\mbox{$N(\mathrm{N^+})\approx 1.5\times 10^{17}$~cm$^{-2}$} towards both sources. 
This is in agreement with 
recent \emph{Herschel}-HIFI observations of [C~{\sc II}] 158~$\mu$m, also observed in 
absorption in the same sight-lines,  if $\approx7-10$~\% of  all C$^+$ 
ions exist in the WIM on average.   
Using an  abundance ratio of  $\mathrm{[N]/[H]} = 6.76\times10^{-5}$ in the gas phase we find that the 
mean electron and proton volume densities are  $\sim0.1-0.3$~cm$^{-3}$     
assuming a WIM  volume filling fraction of $0.1-0.4$  with a corresponding line-of-sight filling
fraction of $0.46-0.74$. 
A low density    and a high WIM filling fraction 
are also 
supported by {\tt RADEX} modelling of the [N~{\sc II}] 205~$\mu$m absorption and emission together with visible emission lines attributed mainly to the WIM. 
The detection of the 205~$\mu$m line in absorption   emphasises the importance 
of  a high spectral resolution, and also offers a new tool for investigation of the WIM.  
}

   \maketitle
%

\section{Introduction}

Among the brightest far-infrared spectral lines in the Galaxy are the forbidden transitions 
of singly ionised nitrogen  
at  122 and 205~$\mu$m  providing   useful probes of 
the Galactic large-scale and spiral 
structure.     
These transitions   are  unfortunately generally  inaccessible to  ground based facilities 
because of  the opaque   atmosphere at these frequencies.    
The
205~$\mu$m transition of N$^+$ was first detected using FIRAS onboard
COBE \citep{1991ApJ...381..200W, 1999ApJ...526..207F}. 
FIRAS performed the
first nearly all-sky far-infrared survey of the Galactic emission at
wavelengths between $\sim$100~$\mu$m and 1~cm and detected the strong and
widespread emission from the forbidden transitions [C~{\sc II}] 158~$\mu$m
and [N~{\sc II}] 205~$\mu$m, although with low spatial and spectral
resolution.
  Since the ionisation potential of nitrogen is 14.53~eV, the
[N~{\sc II}] emission originates in the ionised gas in localised  H~{\sc II} regions and in 
 the widespread  warm
and low-density ionised interstellar gas \citep[WIM; review by][]{2009RvMP...81..969H}. 
This is in contrast to the bulk of ionised carbon emission which   traces the 
cold neutral medium (CNM). 
While the [N~{\sc II}] emission
from a discrete source was first detected using the Kuiper Airborne
Observatory towards the H~{\sc II} region G333.6$-$0.2 
\citep{1993ApJ...413..237C}, more
recently [N~{\sc II}]  emission from the Carina Nebula was detected 
for the first time using a
ground-based telescope  
\citet[][SPIFI on AST/RO at the South Pole]{2006ApJ...652L.125O, 2011ApJ...739..100O}.

 \begin{table*}[\!htb] 
\centering
\caption{Source sample: properties and resulting continuum intensities and noise levels.}
\begin{tabular} {lccccccc } 
 \hline\hline
     \noalign{\smallskip}
Source  & R.A.  & Dec. & Distance & $\varv_\mathrm{LSR}$\tablefootmark{a} & l.o.s\tablefootmark{b}  &
$T_\mathrm{C}$\tablefootmark{c}  & $1\sigma$/$T_\mathrm{C}$\tablefootmark{d}\\ 
 &  ($J$2000)  &  ($J$2000)  & (kpc) & (km~s$^{-1}$)  & (km~s$^{-1}$) & (K)\\
     \noalign{\smallskip}
     \hline
\noalign{\smallskip}  
W31C (G10.6$-$0.4)  & 18:10:28.7 & -19:55:50  & 4.95\tablefootmark{e} & $-$3 & 10--61 & 5.6  & 0.015\\
W49N (G43.2$-$0.1) & 19:10:13.2&  +09:06:12  & 11.1\tablefootmark{f}  & +11 & 20--84  & 7.0  & 0.013\\
W51  (G49.5$-$0.4) & 19:23:43.9&  +14:30:30.5 & 5.4\tablefootmark{g} & +57 & 1--45    &  6.7   &  0.011\\
G34.3+0.1 & 18:53:18.7& +01:14:58 & 3.3\tablefootmark{h} & +58 & 8--45  & 5.6   &  0.014\\
    \noalign{\smallskip} \noalign{\smallskip}
\hline 
\label{Table: sources}
\end{tabular}
\tablefoot{
\tablefoottext{a}{Source LSR velocity.} 
\tablefoottext{b}{LSR velocity range of foreground absorbing gas.} 
\tablefoottext{c}{Single sideband (SSB) continuum intensity as measured in the Dual Beam Switch data. } 
\tablefoottext{d}{Rms noise  at a resolution of 1~km~s$^{-1}$  divided by $T_\mathrm{C}$.} 
\tablefoottext{e}{\citet{2014ApJ...781..108S}.} 
\tablefoottext{f}{\citet{2013ApJ...775...79Z}.}
\tablefoottext{g}{\citet{2010ApJ...720.1055S}.}
\tablefoottext{h}{\citet{1994ApJ...436..117K}.}
} 
\end{table*}

Fine-structure levels within the ground term of N$^+$ will be excited mainly by 
electrons in the WIM and nebular regions and by both electrons and H atoms in 
the mostly neutral  CNM    and warm neutral medium. 
The rates of electron collisions at kinetic temperature $T_\mathrm{K}\sim8\,000$~K compete 
with radiative rates when the electron density is $\sim40$~cm$^{-3}$  while the 
corresponding neutral density would need to exceed 100~cm$^{-3}$. Thus in 
gas at lower densities, like the WIM at $n_\mathrm{e}<1$~cm$^{-3}$, the [N~{\sc II}]
excitation will remain very subthermal,    
and since the volume filling fraction of the WIM is estimated to be \mbox{$\sim0.1-0.4$} 
 \citep{2009RvMP...81..969H},   
extended regions in the Galaxy 
will    be difficult to probe with [N~{\sc II}]  emission  
 without sufficient
sensitivity. 
However, the diffuse interstellar gas 
with   little or no excitation can be probed with high sensitivity in absorption
along sight-lines towards bright far-infrared continuum sources.
Therefore, the diffuse ionised nitrogen, which has
previously only been observed in emission, can be detected in 
absorption towards a strong continuum background source  at   high
spectral resolution.
The lines need to be
spectrally resolved in order to 
prevent blending of the weak foreground absorption features
with the stronger background emission which may even lead to partial or
complete disappearance of the background emission  by the  
foreground absorption.  
The sensitive 
Heterodyne Instrument for the Far-Infrared (HIFI) onboard the \emph{Herschel} Space Telescope, 
designed to 
perform very high  spectral resolution  observations at THz frequencies  (\mbox{0.48--1.25~THz} and \mbox{1.41--1.91~THz}),
enabled such observations  of the [N~{\sc II}] 205~$\mu$m line.   
In the framework of 
the \emph{Herschel} 
\citep{Pilbratt2010, 2012A&A...537A..17R}
key programme 
PRISMAS\footnote{{\tt http://astro.ens.fr/?PRISMAS}} (PRobing InterStellar Molecules with Absorption line Studies)  
absorption in numerous molecular and atomic lines has been studied 
towards a sample of massive star forming regions in the 
Galactic plane  \citep[e.g.][]{2010A&A...518L.110G, 2010A&A...521L..10N, 2010A&A...521L..13M, 2010A&A...521L..15F, 2012A&A...543A.145P}.

In this paper, we present \emph{Herschel}-HIFI observations of   the  [N~{\sc II}]  205~$\mu$m   transition towards 
W31C, W49N, W51, and G34.3+0.1. 
The data   are a part of the OT1 programme 
\emph{Diffuse ISM phases in the inner Galaxy} (PI Maryvonne Gerin) in which the 
fine structure lines of ionised nitrogen and carbon  and the ground  and first excited states  
of neutral carbon were observed with the goal to   characterise the diffuse neutral and ionised interstellar medium targeting the PRISMAS sources     
\citep[e.g.][]{2010Godard, 2011A&A...525A.116G, 2013ApJ...762...11F}.  
The first results   of  the ionised and neutral carbon   observations    have recently 
been presented in \citet{2012RSPTA.370.5174G, 2014Gerin}.

 \section{Observations and data reduction}

The observations of 
the [N~{\sc II}] fine  structure transition \mbox{$^3P_1 - ^3P_0$}   
at 205.178~$\mu$m   \citep[1\,461.13190~GHz;][]{1994ApJ...428L..37B} 
are summarised in Table~\ref{Table: obsid} 
(on-line material). 
We note that the 205~$\mu$m transition has three hyperfine structure components 
where the second strongest component lies 1.5~km~s$^{-1}$ above the  main component   
causing  a slight broadening of the line profile.
The targeted sources and their properties are listed in Table~\ref{Table: sources}.

Because of the extended 
N$^+$ emission across the Galaxy 
we    
used the load chop   mode in which an internal cold load is used as  reference.  
In this way possible  
contamination from  weak  emission in the off beam was minimised. 
This was complemented with observations in 
dual beam switching (DBS) fast chop mode  
allowing an even better measurement of the continuum than with the load chop mode. The 
reference beams    in these observations were located within 3\arcmin~on either side of the source. 

We  used  the upper sideband of band 6a and  
the wide band spectrometer (WBS) with a bandwidth of 
4$\times$1~GHz and an effective spectral  
resolution of 1.1~MHz  
($\Delta \varv = 0.23$~km~s$^{-1}$).    
The   half-power beam width of the telescope is  
15.0\arcsec~at   1\,410\,GHz 
and the   
calibration uncertainty is $\lesssim$13\% for band 
6a\footnote{\tt http://herschel.esac.esa.int/Docs/HIFI/html/ch5.html}.

The data were processed using the standard \emph{Herschel} Interactive Processing Environment\footnote{
{\tt http://herschel.esac.esa.int/HIPE\_download.shtml}},
version 11.1,    up to  level 2, providing fully calibrated  double side band  spectra    in 
the $T_\mathrm{A}^*$  antenna temperature   intensity scale where the lines are 
calibrated on a single side band (SSB) scale. Because HIFI is
intrinsically a double side band  instrument, the continuum has to be
divided by two to be properly scaled. 
The \emph{FitHifiFringe} task was used to fit and remove standing waves from the spectra.
We also   used the task   
\emph{hebCorrection} in HIPE 12.1 as a second approach to remove the  standing waves
which gave very similar results to those obtained  with  \emph{FitHifiFringe}. 
The FITS  files were   exported to  the spectral line analysis   software packages  
{\tt xs}\footnote{
{\tt http://www.chalmers.se/rss/oso-en/observations/\\data-reduction-software}} and 
{\tt CLASS}\footnote{
{\tt https://www.iram.fr/IRAMFR/GILDAS/doc/html/\\class-html/class.html}}
which were  used in parallel in the subsequent data reduction.  
Both polarisations were included in the averaged noise 
weighted  spectra, and for the DBS data   both LO-settings were also included. The resulting averages    
were   convolved to a resolution of 1~kms$^{-1}$.   
All spectra shown in the figures and used in the analysis are load chop data with continuum levels   
scaled to   $T_\mathrm{C}$  obtained from the  DBS measurements.
 The DBS spectra were also used to check emission in the off-beam in 
Sect.~\ref{section: results}.

\section{Results}\label{section: results}

\begin{figure}
\begin{center}
\resizebox{\hsize}{!}{ 
\includegraphics{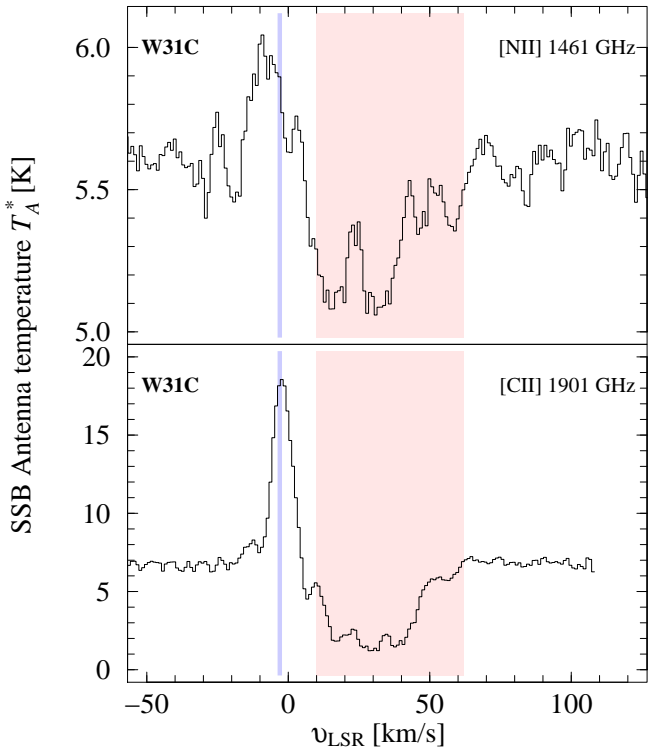}} 
\caption{Single sideband WBS spectra of [N~{\sc II}]~205~$\mu$m and [C~{\sc II}]~158~$\mu$m 
towards W31C.  
The $\varv_\mathrm{LSR}$ of the H~{\sc II} region is marked in blue and the velocity of the line-of-sight 
gas is marked in red.
Absorption of N$^+$   traces the 
warm ionised medium,  while  the C$^+$ absorption  
mainly traces the cold neutral medium, and, at   $\approx 7-10$~\% level also the WIM. 
}      
\label{fig_w31c} 
\end{center}
\end{figure}

\begin{figure}
\begin{center}
\resizebox{\hsize}{!}{ 
\includegraphics{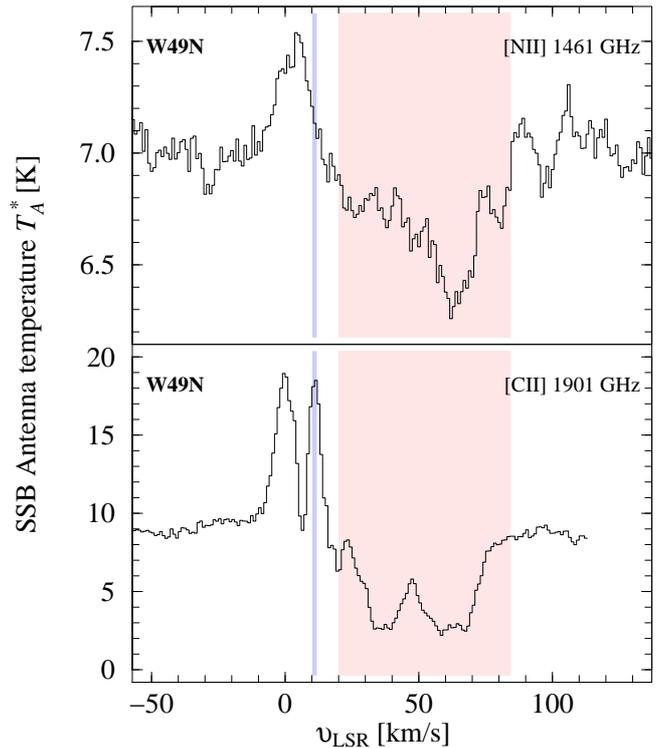}} 
\caption{W49N. Notation  as in Fig.~\ref{fig_w31c}.}   
\label{fig_w49n} 
\end{center}
\end{figure}

\begin{figure}
\begin{center}
\resizebox{\hsize}{!}{ 
\includegraphics{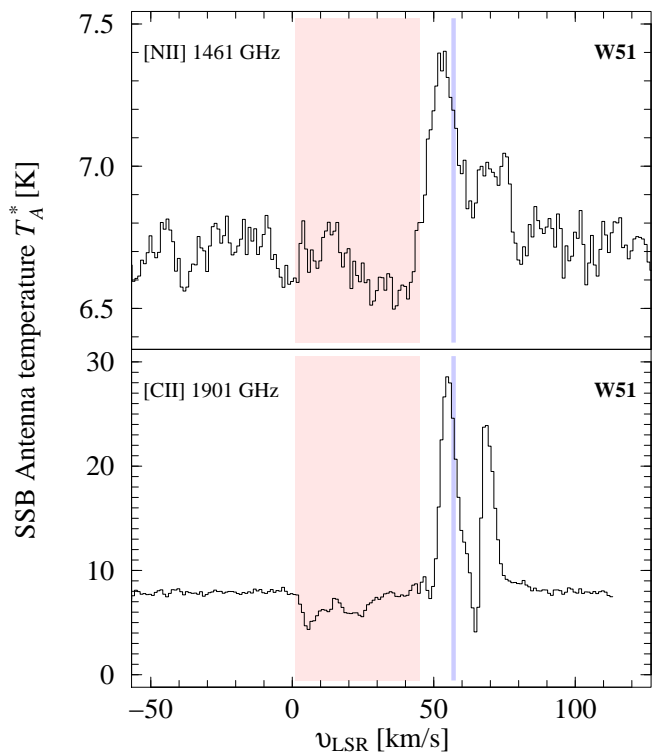}}  
\caption{W51. Notation  as in Fig.~\ref{fig_w31c}.}      
\label{fig_w51}
\end{center}
\end{figure}

\begin{figure}
\begin{center}
\resizebox{\hsize}{!}{ 
\includegraphics{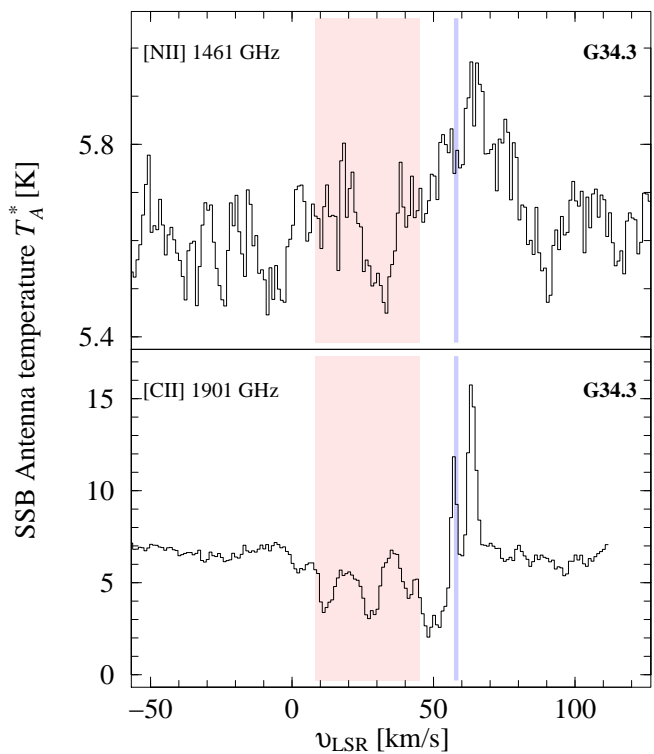}}  
\caption{G34.3+0.1. Notation  as in Fig.~\ref{fig_w31c}.}      
\label{fig_g34}
\end{center}
\end{figure}

Figures~\ref{fig_w31c}--\ref{fig_g34} show the resulting   [N~{\sc II}] 205~$\mu$m data   
together with comparison spectra of the  [C~{\sc II}] 158~$\mu$m line \citep{2012RSPTA.370.5174G,2014Gerin}.  
We find [N~{\sc II}] emission     at the source velocities 
towards all four sources even though the 
emission from G34.3+0.1  is weak. In addition, we  clearly 
detect, for the first time, ionised nitrogen in  
\emph{absorption}   towards W31C and W49N from low-density 
foreground material.  
A  hint of
absorption in the sight-line  towards W51 is also seen 
at $\varv_\mathrm{LSR}\sim20-40$~km~s$^{-1}$.

The optical depths per unit velocity interval, $\tau_\nu$, are derived from  
\begin{equation}\label{antenna temp}
T_\mathrm{A}^* = T_\mathrm{C}\,e^{-\tau_\nu} + J(T_\mathrm{ex})\,(1-e^{-\tau_\nu})\ ,
\end{equation} 
where  
$J(T_\mathrm{ex}) = h\nu_\mathrm{ul}/k \times(\exp(h\nu_\mathrm{ul}/k\,T_\mathrm{ex})-1)^{-1}$, 
$\nu_\mathrm{ul}$ is the frequency 
of the transition, and
$T_\mathrm{ex}$   the excitation temperature. We have here   neglected  the small contribution from the cosmic microwave background radiation and the Galactic radiation field. 
Assuming that  \mbox{$J(T_\mathrm{ex})\!\ll\!T_\mathrm{C}$} (or $T_\mathrm{ex}\!\ll\!27$~K)
we estimate the line opacities  
as \mbox{$\tau_\nu\!=\!-\ln{(T_\mathrm{A}^*/T_\mathrm{C}})$}. 
The total integrated opacities  
are   obtained by summing over the line-of-sight velocities listed in 
 Table~\ref{Table: sources}.  
For   W51 and G34.3+0.1 we estimate   upper limits using $3\sigma$ noise levels. 
The resulting integrated opacities, $\int \tau_\nu \mathrm{d}\varv$, are 3.2, 3.3, $\lesssim\!1.6$, and $\lesssim\!1.5$~km~s$^{-1}$ for 
W31C, W49N, W51,  
and G34.3+0.1, respectively.

The N$^+$ column densities in the foreground gas are estimated  using    the relation 
\begin{equation}\label{N_lower}
N(\mathrm{N}^+) = 8\pi  \frac{\nu_\mathrm{ul}^3}{c^3}   \frac{g_\mathrm{l}}{g_\mathrm{u}\,A_\mathrm{ul}}  \int \tau_\nu \,\mathrm{d}\varv =  4.7 \times 10^{16} \int \tau_\nu \,\mathrm{d}\varv \, \, \, \mathrm{[cm^{-2}]}\ ,  
\end{equation} 
assuming that \mbox{$J(T_\mathrm{ex}) \!\ll\!h \nu_\mathrm{ul}/k=70$~K} and that all N$^+$ ions are in 
the ground state. 
Resulting column densities 
  and upper limits  
are found in Table~\ref{Table: columns}. 

In order to constrain  the  excitation temperature   
of the foreground gas towards W31C and W49N 
we compare the observed 
absorption line spectra   with  spectra taken towards positions 
close to the observed lines of sight, but offset from the background continuum. 
Such  spectra could potentially show  N~{\sc II} emission. 
Equation~(\ref{antenna temp}) then becomes 
 \begin{equation}\label{OFF spectra}
T_\mathrm{A}^*(\mathrm{OFF}) =  J(T_\mathrm{ex}) \,(1-e^{-\tau_\nu})\ .  
\end{equation}   
We obtained the comparison OFF spectra by taking the difference of the load  chop and DBS spectra to 
produce a mean spectrum of the OFF positions approximately 3\arcmin~from the line of sight.  
We do not detect any emission at the
velocities corresponding to the foreground features  in any of the OFF spectra, but we do find 
emission from the H~{\sc II} regions at the source velocities not affecting the line-of-sight absorptions.   
We thus  use the measured rms noise levels, $\sim$0.12~K  at a resolution of 1~km~s$^{-1}$,  
 as   limits for 
$T_\mathrm{A}^*(\mathrm{OFF})$ in the sight-lines together with  the corresponding maximum opacity, 
$\tau_\mathrm{max}\approx 0.11$.   
In this way, 
 we find  $T_\mathrm{ex}\lesssim17$~K  as an upper limit  for the excitation
temperature of N$^+$ in the foreground gas. 
We note that this procedure assumes  that the same component is responsible for 
both the emission and absorption.
 Some N$^+$ can also exist   in the neutral medium (see Sect.~\ref{section: discussion}),  
but here the N$^+$ abundances are expected to be several orders of magnitudes 
lower than those in  the WIM.

\begin{table}[\!ht] 
\centering
\caption{
Resulting  column densities, $N$(N$^+$),  
column density ratios with the related species  C$^+$   
in the diffuse line-of-sight gas, and average ionised hydrogen volume density in the WIM\tablefootmark{a}.
}
\begin{tabular} {lccc} 
 \hline\hline
     \noalign{\smallskip}
Source  & $N$(N$^+$)  & $N$(C$^+$)/\,$N$(N$^+$)\tablefootmark{b}   
& $n_\mathrm{H^+}$\tablefootmark{c}\\   
 \noalign{\smallskip}
  	& (cm$^{-2}$) &&   (cm$^{-3}$)\\
     \noalign{\smallskip}
     \hline
     \noalign{\smallskip}
W31C & 1.5$\times$10$^{17}$	 
&  41    
&  $0.20-0.32$  \\ 

W49N & 1.6$\times$10$^{17}$	       
& 43        
&  $0.09-0.15$    \\
 
W51  & $\lesssim7\times10^{16}$   
& $\gtrsim15$   
& $\lesssim0.09-0.14$      \\

G34.3+0.1 &  $\lesssim7\times10^{16}$   
& $\gtrsim22$   
&  $\lesssim0.14-0.23$   \\ 
 
  \noalign{\smallskip}
\hline 
\label{Table: columns}
\end{tabular}
\tablefoot{
\tablefoottext{a}{Estimated over the velocity ranges listed in Table~\ref{Table: sources}.
} 
\tablefoottext{b}{$N$(C$^+$) from \citet{2012RSPTA.370.5174G, 2014Gerin}.} 
\tablefoottext{c}{Assuming a volume filling fraction  $0.1-0.4$  (corresponding to a line-of-sight filling fraction
 \mbox{$f=0.46-0.74$} in Eq.~\ref{Hp column}  obtained from the volume filling factor raised to the 1/3 power).} 
}
\end{table}

\section{Discussion} \label{section: discussion}
Since the ionisation potential of carbon is 11.3~eV the C$^+$ ion  largely traces the CNM, 
and  to a lesser extent     the WIM. The fraction of   C$^+$ ions existing in the WIM  
 can be estimated from a comparison of the C$^+$ and
N$^+$ column densities since  
the [N~{\sc II}]~205~$\mu$m    and the  [C~{\sc II}]~158~$\mu$m   lines  
  have nearly identical critical densities, hence 
their line ratio in the WIM is   only  a function of the [C]/[N]  gas phase abundance  ratio. 
We consider two cases: (\emph{i})      
$\mathrm{[C]/[N]} = 3.2$  \citep{2007ApJ...654..955J, 2004ApJ...605..272S} which includes depletion 
in translucent and diffuse gas, 
 and    
(\emph{ii}) $\mathrm{[C]/[N]} = 4.0$ using solar elemental abundances without depletion 
\citep{2009ARA&A..47..481A}. A comparison of these ratios   with the     
  derived $N(\mathrm{C}^+)/N(\mathrm{N}^+)$ ratios  
listed in Table~\ref{Table: columns} 
suggests 
that on average $\approx7-10$~\% of all C$^+$ ions exist    in the WIM,
 in agreement with 
 \citet{2014Gerin}. 

The fully ionised WIM  allows an estimate of its 
column and average volume densities of  ionised hydrogen  via 
\begin{equation}\label{Hp column}
N(\mathrm{H^+}) =  N(\mathrm{N^+})\times (\mathrm{[N]/[H]})^{-1} \approx f \, n_\mathrm{H^+}\,  s\ \   [\mathrm{cm^{-2}}]\ , 
\end{equation}
where $f$ is the filling factor of the WIM along the line of sight, and $ s$ is the distance  
to the source. 
We take for reference an  abundance  ratio $\mathrm{[N]/[H]} = 6.76\times 10^{-5}$ 
 in the gas phase \citep{2009ARA&A..47..481A}. 
Since  $f$ cannot be higher than unity,   the lower limits of the average 
$n_\mathrm{H^+}$   towards W31C and W49N are 
$\sim0.15$ and $\sim0.07$~cm$^{-3}$, respectively.  
Using more realistic values of the  
filling factor   we find  an average  $n_\mathrm{H^+} \sim0.1-0.3$~cm$^{-3}$    (Table~\ref{Table: columns}).

Emission lines of [N~{\sc II}]  are taken to trace ionised hydrogen 
and thus the rate of formation of massive stars that produce ionising 
photons in the Milky Way and in high-redshift starburst galaxies 
\citep[e.g.][]{2012ApJ...752....2D}. \citet{2013ApJ...765L..13Z} used 
the low spectral resolution SPIRE instrument onboard 
\emph{Herschel}  to 
survey [N~{\sc II}] 205~$\mu$m line emission in a sample of 70 local 
luminous infrared galaxies and found both low line to continuum flux ratios 
in the most luminous systems and an unexplained scatter in that ratio over the whole sample. 
Our detection of  the 205~$\mu$m line in absorption 
from low-density foreground gas 
highlights the need for high spectral resolution in investigating
 N$^+$ column densities in galaxies  since absorption  of the [N~{\sc II}] line in 
low-density gas can affect the total integrated intensity of the line
emission.

\subsection{Radex modelling}

Besides   the 
emission lines in the submm at 205 and 122~$\mu$m, interstellar atomic nitrogen ions    have
also been observed at   visible  wavelengths 
(6\,583, 6\,548, and 5\,755~\AA), and 
through absorption lines in the far ultraviolet (1\,084~\AA). 
All these  lines are thought to originate in a combination of localised photoionised nebulae and the widespread WIM. 
To demonstrate that our observed submm-wave 
absorption is quantitatively consistent with the emission lines attributed to   the WIM we constructed
a simple model    using    
the non-LTE radiative transfer code  
{\tt RADEX}\footnote{\tt{http://www.sron.rug.nl/$\sim$vdtak/radex/radex.php}} \citep{2007A&A...468..627V}.

In order 
to describe the observable  intensities and optical depths of the N$^+$ lines, we adopt 
term energies and transition probabilities from the 
NIST database\footnote{{\tt http://www.nist.gov/pml/data/asd.cfm} is the 
Atomic Spectra Database version 5.1 maintained 
by the U.S. National Institute of Standards and Technology: Kramida, Ralchenko, \& 
Reader et al. (2013)}. Electron-impact collision strengths and additional transition probabilities 
are taken from \citet{2011ApJS..195...12T}. Neutral-impact collision rates for the excitation 
of the ground-term fine-structure levels are not known; therefore, the corresponding rates 
for the isoelectronic neutral system \mbox{H + C($^3$P)} computed by \citet{2007ApJ...654.1171A}  
 have been scaled upwards by a factor of 5 for the ion-neutral system \mbox{H + N$^+ (^3{\rm P})$}. 

We computed the spectrum of a 58-level N$^+$ ion from submm to EUV wavelengths  
with {\tt RADEX} and  the atomic data described above and also  
estimated the  H$\alpha$ intensity 
  for  Case~B recombination \citep{1987MNRAS.224..801H}  for an assumed gas phase 
nitrogen abundance of \mbox{$\mathrm{[N]/[H]}=6.76\times 10^{-5}$}. 
The model parameters are consistent with chemical modelling where 
the N$^+$ ion is  produced by cosmic ray ionisation of N and by charge transfer, 
\mbox{H$^+$ + N $\to$ H + N$^+$}, \citep{2005PhRvA..71f2708L}. The removal can be  
by direct radiative recombination and by dielectronic recombination 
\citep{1983A&A...126...75N,2004A&A...417.1173Z, 2006ApJS..167..334B,2009ApJ...694..286B}  
and by the reverse of the charge transfer 
reaction. In molecular regions N$^+$ is also destroyed by reaction with H$_2$, for which 
we adopt the rate coefficient recommended by \citet{2012ApJS..199...21W}.  The cosmic 
ray ionisation rate of hydrogen is taken to be $\zeta_0 = 2\times 10^{-16}$~s$^{-1}$ 
\citep[see][]{2012ApJ...758...83I} and that for nitrogen is $2.1 \zeta_0$. We assume 
that the abundance of N$^+$ is governed by the rates of these formation and destruction 
processes in steady state. The abundance of N$^+$ then depends only on the total density 
of hydrogen nuclei $n_{\rm H}$, the kinetic temperature of the gas $T_\mathrm{K}$, and the fractional 
abundances of electrons, $f_\mathrm{e}$, and hydrogen molecules, $f_{{\rm H}_2}$. 
 
Models of N$^+$ in the WIM 
at $T_\mathrm{K} = 8\,000$~K 
reproduce both the submm-wave absorption towards W49N and W31C reported here and the intensities 
of [N~{\sc II}] 6583,  5755, and H$\alpha\; 6563$~\AA\ reported by 
\citet{2001ApJ...548L.221R}. The latter refer to emission lines over a one-degree region in 
the direction G$130.0-7.5$ in several broad velocity components with corresponding kinematical 
distances between 0.6~kpc and 9.7~kpc. 
We estimate that the effect of dust extinction along the sight-lines is  small 
compared with variations 
in (\emph{i}) the total nitrogen abundance over 10~kpc sight-lines across the Galaxy, 
and  (\emph{ii}) variations in electron density and temperature.  
Models of column density 
\mbox{$N({\rm N}^+) = 1.5\times 10^{17}$~cm$^{-2}$}  
over a total linewidth $\Delta \varv=50$~km~s$^{-1}$, in agreement with our results in 
Sect.~\ref{section: results}, with uniform densities $n_{\rm H} = 0.05-0.1$~cm$^{-3}$ and electron fractions  $f_e\sim 1$ yield an integrated 
optical depth $\int \tau\,\mathrm{d}\varv = 3.2$~km~s$^{-1}$ in the 205~$\mu$m  line. The same models 
produce the following intensities in the visible lines: $\Phi(6583) = 10$ to 20~Rayleigh, somewhat 
larger than the  4.3~R observed; $\Phi({\rm H}\alpha) = 10$ to $19$~R, comparable to the observed 10.1~R; 
and $I(5755)/I(6583) = 0.8\%$, close to the observed intensity ratio $(0.96\pm 0.20)\%$.

In these models  
the excitation temperature of the  205~$\mu$m line is $9-10$~K, 
which explains how this easily excited line can appear in absorption towards our submm-wave continuum sources: 
the integrated diffuse emission would be \mbox{$\int T_{\rm RJ}\,\mathrm{d}\varv < 0.2$~K~km~s$^{-1}$}, 
well below our  detection limits. The  model  intensity ratio 
of the two fine-structure lines, $I(122)/I(205) \approx 0.6$ is somewhat smaller than the global 
average $1.1\pm 0.1$ measured with the FIRAS instrument on the COBE  satellite. 
The ratio is, however, 
consistent with that expected for diffuse gas since the 
higher observed ratio was due to higher density
ionised gas closely associated with star forming regions 
excluded from the observations
 \citep{1993ApJ...405..591P}.
Models of the WIM or the  
neutral medium   with $f_\mathrm{e}<0.5$ yield optical intensity ratios $I(6583)/I({\rm H}\alpha)$ that 
are much larger than observed in diffuse emission. Photoionised nebulae with densities 
$n_\mathrm{e} \gg 0.1$~cm$^{-3}$ can be ruled out because the 205~$\mu$m line would appear 
strongly in emission. Models of neutral, partly molecular gas at low temperature, $T_\mathrm{K}\lesssim100$~K, 
are unable to explain the observed absorption because the N$^+$ abundance in neutral gas is too 
low at the adopted ionisation rate and the observed abundance would imply an amount of NH$^+$ in 
conflict with observed upper limits   \citep{2012A&A...543A.145P, 2014Persson}. 
The total cooling provided by [N~{\sc II}] in our models is dominated by the 
 6\,548~\AA and 6\,583~\AA~transitions (86~\%), while the 205~$\mu$m and 122~$\mu$m fine-structure transitions account  for 12~\%.

In summary,  the  {\tt RADEX} modelling of the  205~$\mu$m absorption lines observed over long, $\sim5-10$~kpc, sight-lines towards 
W49N and W31C are compatible with a widespread WIM at $T_\mathrm{K}\sim 8000$~K in which the density 
$n_e \approx n_{\rm H^+} \sim 0.05-0.1$~cm$^{-3}$. The same conditions explain the diffuse 
optical emission lines of [N~{\sc II}] and H$\alpha$ \citep{2001ApJ...548L.221R}. The real 
interstellar medium  is, however,  
more complex than our static, uniform models; in particular, the WIM represents an average 
over space and time of wildly varying conditions \citep{2011ApJ...727...35D}.

\begin{acknowledgements}
HIFI has been designed and built by a consortium of institutes and university departments from across Europe, Canada and the United States under the leadership of SRON Netherlands Institute for Space Research, Groningen, The Netherlands and with major contributions from Germany, France and the US. Consortium members are: Canada: CSA, U.Waterloo; France: CESR, LAB, LERMA, IRAM; Germany: KOSMA, MPIfR, MPS; Ireland, NUI Maynooth; Italy: ASI, IFSI-INAF, Osservatorio Astrofisico di Arcetri-INAF; Netherlands: SRON, TUD; Poland: CAMK, CBK; Spain: Observatorio Astronómico Nacional (IGN), Centro de Astrobiología (CSIC-INTA). Sweden: Chalmers University of Technology - MC2, RSS \& GARD; Onsala Space Observatory; Swedish National Space Board, Stockholm University - Stockholm Observatory; Switzerland: ETH Zurich, FHNW; USA: Caltech, JPL, NHSC.
CMP and JHB acknowledge generous support from the Swedish National Space Board. 
JRG thanks the Spanish MINECO  for funding support under grants CSD2009-00038, AYA2009-07304 and AYA2012-32032.
Thanks also to Paul Goldsmith and the anonymous referee	 whose constructive comments improved the paper. 
\end{acknowledgements}

\bibliographystyle{bibtex/aa}
\bibliography{references}

\Online
\appendix

 \section{Tables}

 \begin{table}[\!htb] 
\centering
\caption{\emph{Herschel} observational identifications (OBSIDs) of the observed [N~{\sc II}] 1\,461~GHz transition in HIFI band 
6a presented in this paper.
}
\begin{tabular} {l  lccc } 
 \hline\hline
     \noalign{\smallskip}
Source  & Obs mode\tablefootmark{a} & Date &	OBSID    \\    \noalign{\smallskip}

&  & (GHz)    \\
     \noalign{\smallskip}
     \hline
\noalign{\smallskip}  

W49N       & DBS-A & 2011 Oct 8 &  1342230294	  \\    
                  & DBS-B &            & 1342230295   \\
               & LC &            &  1342230296   \\
 
\noalign{\smallskip} \noalign{\smallskip} \noalign{\smallskip} 

G10.6-0.4        & DBS-A & 2012 April 9 & 1342244077 	  \\    
                 & DBS-B &            &1342244078   \\
              & LC &            & 1342244079  \\
\noalign{\smallskip}

\noalign{\smallskip} \noalign{\smallskip} \noalign{\smallskip} 
W51      & DBS-A & 2011 Oct 8 & 1342230299	  \\    
                 & DBS-B &            & 1342230298   \\
                & LC &            &  1342230297   \\
 
\noalign{\smallskip} \noalign{\smallskip} \noalign{\smallskip} 

G34.3+0.1     & DBS-A & 2011 Oct 8 & 1342230291 	  \\    
                & DBS-B &            & 1342230292   \\
                 & LC &            &  1342230293   \\
  
    \noalign{\smallskip}
\hline 
\label{Table: obsid}
\end{tabular}
\tablefoot{
\tablefoottext{a}{Two different frequency settings of the LO were 
performed in dual beam switching (DBS), fast chop, in order to
determine the sideband origin of the signals and to obtain the best
possible continuum measurement. In addition,   load
chop (LC) observations were also performed because of the extended emission of the species.}  
}
\end{table} 

\end{document}